\documentclass{moriond}
\usepackage{hyperref}
\usepackage{amsmath,amssymb}

\def\be{\begin{equation}}
\def\ee{\end{equation}}
\def\bea{\begin{eqnarray}}
\def\eea{\end{eqnarray}}

\newcommand{\Photo}{}

\begin{document}
\vspace*{4cm}
\title{Dark Matter scattering in low threshold detectors¶}

\author{ Simon Knapen }

\address{Lawrence Berkeley National Laboratory, Berkeley, United States, \\
Berkeley Center for Theoretical Physics, UC Berkeley, Berkeley, United States}

\maketitle
\abstracts{
The scattering of sub-GeV dark matter in direct detection experiments happens at characteristic wavelengths comparable or larger than the interparticle spacing. Collective effects in the target material must therefore be accounted for when calculating the scattering rate. For dark matter-nucleon couplings, this implies matching onto the appropriate phonon effective theory and calculating single and multi-phonon scattering amplitudes. For dark matter-electron couplings, we make use of the energy loss formalism to predict the scattering rate. Combining both techniques allows us to derive a formula for the Migdal effect in crystals, which differs from prior calculations performed in atomic systems.
}

\section{Introduction}

The experimental landscape of dark matter detectors has now evolved well beyond the traditional WIMP detectors, by pushing towards ever lower thresholds. Already experiments are taking data which are sensitive the excitation of a single electron-hole pair in a silicon or germanium target.  In addition phonon sensors are being developed which will soon be sensitive to sub-eV energy depositions. These very low thresholds enable new searches for dark matter with a mass well below 1 GeV. 

Dark matter this light posses additional theoretical challenges when modeling the scattering rates in the experiment: the deBroglie wavelength of dark matter with $m_\chi \lesssim 1$ GeV is not negligibly small compared to the typical length scales of the target material. This means that descriptions in terms of a ``billiard ball'' nuclear recoil or a recoil of a free electron are not appropriate. Instead we must account for the long distance structure and interactions present in the target material, which qualitatively impact the scattering rate. For dark matter - nucleon couplings, this implies thinking in terms collective phonon degrees of freedom. For electron recoils we must account for the delocalized nature of the valence electrons, as well as for the long-range screening effects present in all semi-conductors and metals.

%

\section{Phonons\label{sec:phonons}}

For light dark matter with $m_\chi\lesssim 10$ MeV, the deBroglie wavelength $\sim 1/(v m_\chi)$ exceeds the spacing between individual atoms in a typical crystal. It is therefore impossible for the dark matter to recoil against any one specific nucleus. Instead, the dark matter collision will produce one or more collective excitations, i.e.~phonons. To perform the scattering rate calculation, we must effectively integrate out the atoms themselves and transition to an effective theory with only phonon degrees of freedom. This implies performing the corresponding matching calculation for specific dark matter models, to obtain the dark matter-phonon effective operators \cite{Knapen:2017ekk,Griffin:2018bjn}. 

\begin{figure}
\centerline{\includegraphics[height=3cm]{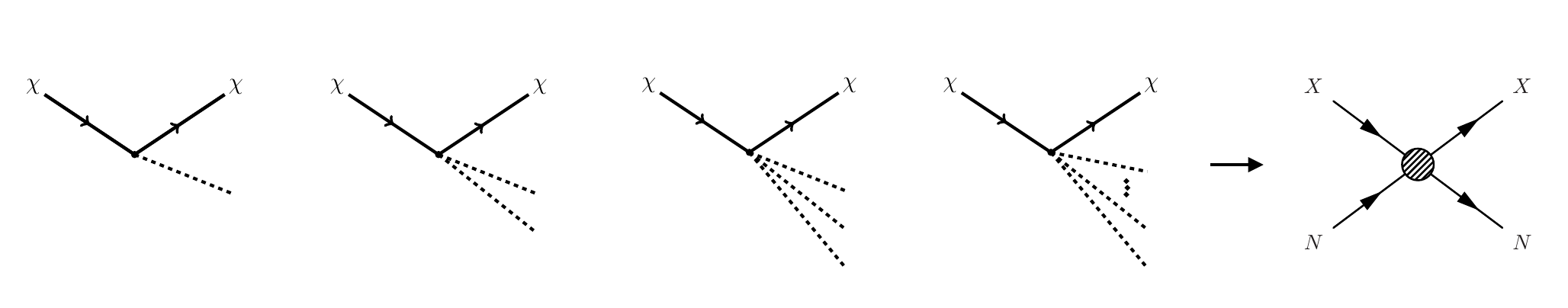}}
\caption[]{From left to right, diagrams representing the LO, NLO, $\mathrm{NNLO}$ and $\mathrm{N^nLO}$ contributions in the $q/\sqrt{2 m_N \omega}$ expansion. In the large $q$ limit, the inclusive cross section converges to the nuclear recoil calculation, represented by the right-most diagram.}
\label{fig:diagrams}
\end{figure}

One of the defining features of the phonon EFT is that it is a derivative expansion in the smallness of the momentum transfer $q$, not unlike pion scattering in chiral perturbation theory. More specifically, we expand in $q/\sqrt{2 m_N \omega}$ with $m_N$ and $\omega$ respectively the nucleus mass and the energy transfer. Each additional phonon we add to the amplitude comes with an additional power of $q/\sqrt{2 m_N \omega}$, such that we can also think of the small $q$ expansion as an expansion in the number of final state phonons, again analogous to chiral perturbation theory \cite{Campbell-Deem:2022fqm}. \footnote{This scaling is however modified strongly by interference effects for the single phonon amplitude, where additional care is needed. \cite{Knapen:2017ekk,Griffin:2018bjn}} For \mbox{$m_\chi\lesssim 1$ MeV} one finds $q\ll \sqrt{2 m_N \omega}$ and the single phonon rate always dominates (left diagram in Fig.~\ref{fig:diagrams}), provided that the experimental threshold is low enough to resolve it. For higher thresholds and higher $m_\chi$, it is not sufficient to only include the single phonon terms and one must go to higher orders in the multiphonon expansion. We first calculated the next-to-leading order term \cite{Campbell-Deem:2019hdx} and more recently the all-orders contributions \cite{Campbell-Deem:2022fqm}. The latter was possible by neglecting the interference between the atoms within the same unit cell, which is justified for momenta larger than the size of the Brillouin zone. The resulting inclusive cross section is proportional to
\begin{equation}
\frac{d \sigma}{d^3 {\bf q} d\omega} \sim \sum_{d}^{\mathfrak{n}} A_d^2   e^{-2 W_d(\mathbf{q})}\sum_n   \bigg(\frac{q^2}{2 m_d} \bigg)^n  \frac{1}{n!} \bigg(\prod_{i=1}^n \int d \omega_{i} \frac{D_d(\omega_i)}{\omega_i} \bigg) \delta \big(\sum_j \omega_j - \omega \big).
\end{equation}
with $A_d$ the mass number of the target atom indexed by $d$, $W_d$ the Debeye-Waller factor, $n$ the number of final state phonons and $D_d$ the partial density of states associated with the atom indexed by $d$. One can show that for large $q$, this formula asymptotes to a gaussian of the form
\begin{equation}
\frac{d \sigma}{d^3 {\bf q} d\omega} \sim \sum_d^{\mathfrak{n}} A_d^2 \sqrt{\frac{2 \pi}{\Delta_d^2}} e^{- \frac{ \big(\omega - \frac{q^2}{2m_d} \big)^2}{2 \Delta_d^2}}
\end{equation}
where the width $\Delta_d$ is an integral over the partial density of states $D_d(\omega)$. At even larger $q$, this gaussian moreover narrows and converges to 
\begin{equation}
\frac{d \sigma}{d^3 {\bf q} d\omega} \sim \sum_d^{\mathfrak{n}} A_d^2 \times \delta\big(  \omega - \frac{q^2}{2m_d} \big)
\end{equation}
which is the $\delta$-function associated with energy-momentum conservation in a classical nuclear recoil. We thus found a single, unifying formula which smoothly interpolates from the single phonon regime all the way to the regime where a free nuclear recoil is a good description of the dark matter scattering process.

\section{Electrons\label{sec:electrons}}
The problem of characterizing the dark matter scattering with the valence electrons in a semi-conductor is complicated by the fact that those electrons are delocalized, interacting and experience screening. This can be addressed by carefully calculating the electron excitation matrix elements with density functional theory (DFT) methods, though here we followed an alternative and a conceptually simpler approach. The rate at which dark matter scatters with electrons is directly related to the rate at which the dark matter looses energy by exciting electrons in the crystal \cite{Knapen:2021run,Hochberg:2021pkt}. The latter is proportional to the imaginary part of the inverse dielectric function, $\mathrm{Im}[\frac{-1}{\epsilon(\omega,\mathbf{k})}]$, which is known as the ``energy loss function'' (ELF) in the material science literature. One advantage of the ELF method is that collective screening effects are automatically included in the rate calculation, while the screening must be included by hand in calculations using the matrix element method. The ELF can also be extracted from experimental data, providing an additional cross check.

\begin{figure}
\centerline{\includegraphics[width=\textwidth]{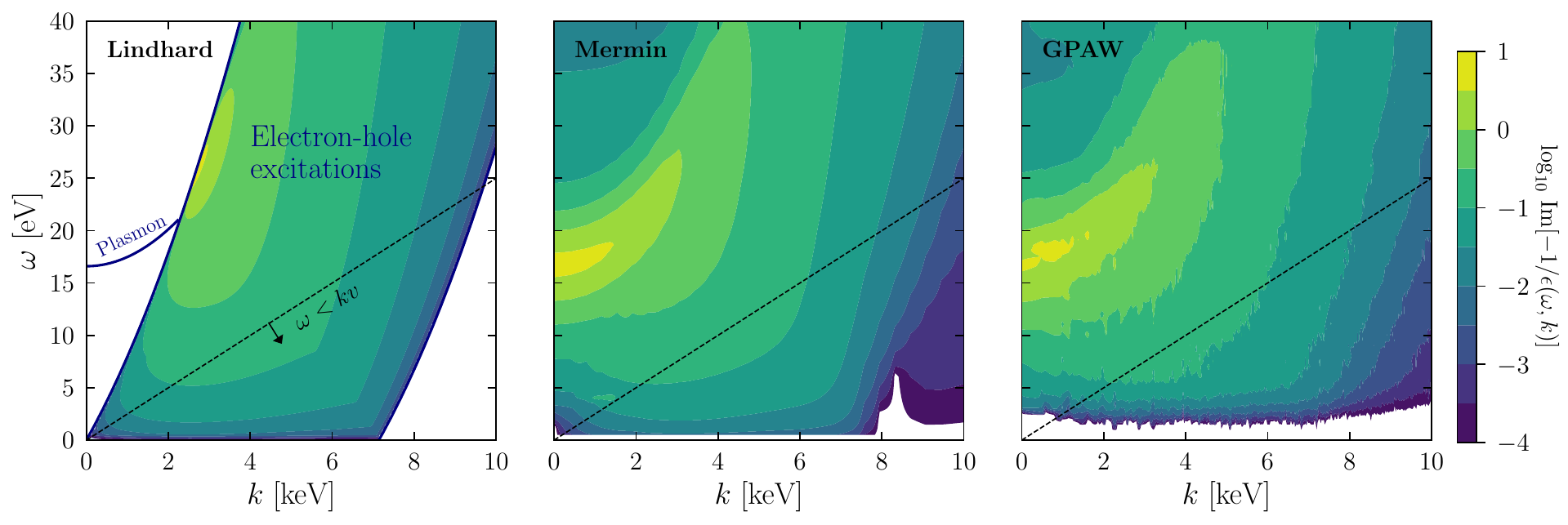}}
\caption[]{Energy loss function, calculated with the Lindhard, Mermin and DFT (GPAW) methods.\cite{Knapen:2021run,Knapen:2021bwg}}
\label{fig:elf}
\end{figure}

We calculated the ELF with three independent methods, as shown in Fig.~\ref{fig:elf}. The Lindhard model is merely a toy model, and describes a non-interacting electron gas without dissipation. The Mermin model is a phenomenological ansatz which has been fit to experimental data from photon absorption measurements. The most sophisticated method is to calculated the ELF directly with DFT methods, for which we used the GPAW code \cite{Enkovaara_2010}. This calculation treats the semi-core electrons as frozen, which leads to a slight underestimation of the rate at high momentum transfers \cite{Griffin:2021znd}. This is being addressed in ongoing work.

\section{The Migdal effect}
\begin{figure}
\centerline{\includegraphics[height=3cm]{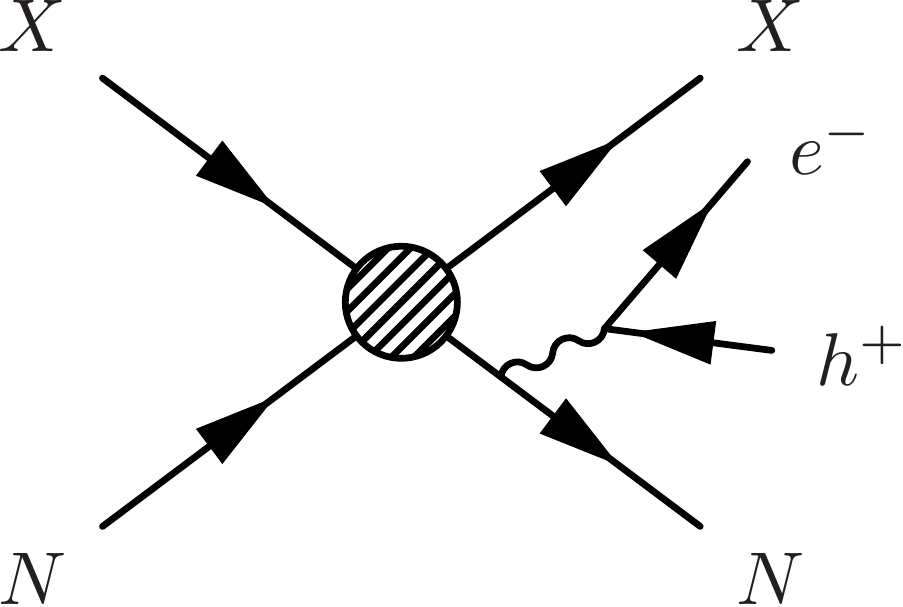}}
\caption[]{Diagrammatic depiction of the Migdal effect: An electron hole pair is created through the electromatic perturbation sourced by the recoiling nucleus.}
\label{fig:migdal}
\end{figure}

There are instances in which it is experimentally desirable to look for electron excitations, as opposed soft nuclear recoils and/or phonon production. If the dark matter predominantly couples to nuclei this is still possible by means of the Migdal effect. The Migdal effect is not a particularly exotic phenomenon: If some amount of energy and momentum is imparted on a nucleus in the target, its induced motion perturbs the electromagnetic potential around it, which in turn may cause the creation of one or more electron hole pairs (see Fig.~\ref{fig:migdal}). In this sense, this process is the condensed matter analogue of the familiar internal conversion process in particle physics, through a virtual photon, e.g.~in $\pi^0 \to \gamma e^+e^-$ etc. 

In a semi-conductor, the main subtlety is that the impact of the surrounding nuclei cannot be neglected: They contribute to the screening of the electromagnetic perturbations and provide a potential in which the recoiling nucleus resides. At very low momenta, the nuclear recoil picture moreover breaks down, as explained in Sec.~\ref{sec:phonons}. We performed a fully self-consistent calculation \cite{Knapen:2020aky}, by combining the multiphonon picture in Sec.~\ref{sec:phonons} with the energy loss function formalism in Sec.~\ref{sec:electrons} (see also \cite{Liang:2020ryg,Berghaus:2022pbu}).

\section{Conclusions}

We provided the first calculations of multiphonon production in dark matter scattering and the Migdal effect in semi-conductors, while accounting for the spectator ions in the crystal. We further improved on the existing dark matter - electron scattering calculations by including screening effects through the energy loss function formalism. All our results are included in our public python package \verb+darkELF+ \cite{Knapen:2021bwg,darkelf}.

\section*{Acknowledgments}

I am delighted to thank Tongyan Lin, Ethan Villarama, Brian Campbell-Deem and Jonathan Kozaczuk for collaboration on the calculations presented in this talk.

\bibliography{moriond}

\end{document}